\documentclass[twocolumn,showpacs,superscriptaddress,prl]{revtex4}
\usepackage{amsmath}
\usepackage{graphicx} 
\usepackage{amssymb}

\begin{document}

\title{Filled Landau levels in neutral quantum gases}

\author{P. \"Ohberg}
\affiliation{Department of Physics, University of Strathclyde,\\
Glasgow G4 0NG, Scotland}
\author{G. Juzeli\={u}nas}
\affiliation{Institute of Theoretical Physics and Astronomy
of Vilnius University,\\
A. Go\v{s}tauto 12, 2600 Vilnius, Lithuania}
\author{J. Ruseckas}
\affiliation{Institute of Theoretical Physics and Astronomy
of Vilnius University,\\
A. Go\v{s}tauto 12, 2600 Vilnius, Lithuania}
\affiliation{Department of Physics, University of Kaiserslautern\\
D-67663 Kaiserslautern, Germany}
\author{M. Fleischhauer}
\affiliation{Department of Physics, University of Kaiserslautern\\
D-67663 Kaiserslautern, Germany}

\begin{abstract}
We consider the signatures of the Integer Quantum Hall Effect in a degenerate
gas of electrically neutral atomic fermions. An effective magnetic field is
achieved by applying two incident light beams with a high orbital angular
momentum. We show how states corresponding to completely filled Landau levels
are obtained and discuss various possibilities to measure the incompressible
nature of the trapped two-dimensional gas. 
\end{abstract}


\maketitle

\section{Introduction}

Recent experimental advances in trapping and cooling atoms has
enabled us to control and engineer the quantum states of delicate quantum gases
such as the degenerate Fermi gases and Bose-Einstein condensates \cite{butts97,demarco99,mewes00,bec_stri}. Ultra-cold atomic gases have turned out to be a
remarkably good medium for studying a wide range of physical phenomena. This is
mainly due to the fact that it is relatively easy to experimentally manipulate
parameters of the system, such as the strength of interaction between the
atoms, properties of a lattice in which the atoms are trapped, the geometry of
an external trap etc. Such a freedom of manipulating the parameters is usually
not accessible in other systems known from condensed matter or solid state
physics. In the present paper we study trapped spin polarised fermions. Using
fermions we naturally have a situation which closely resembles the electronic
case with one important exception: the atoms are electrically
neutral, and there is no vector potential term due to a magnetic field
acting on the atoms. Therefore a direct analogy between the atomic and
electronic case is not necessarily straightforward.

The idea of producing {\it effective} magnetic fields in quantum gases has been
investigated by several authors usually in connection with optical lattices
\cite{jaksch03,anders04,mueller04,levenstein-non-abelian-preprint05}
and external rotation \cite{bretin04,schweikhard04}. It has recently been
realised that in certain situations, in particular in a rotating frame, trapped
atomic quantum gases can be used to recreate the physical state corresponding
to filled Landau levels, and consequently quantum Hall states
\cite{Baym03,fischer03,bretin04,schweikhard04,cooper01,ho01,baranov04}. Present
experimental techniques used for reaching filled Landau levels involve stirring
of a Bose-Einstein condensate \cite{bretin04,schweikhard04}. The ultimate goal
here being to reach the state containing as many vortices in the superfluid as
there are atoms. Experimentally it is however a rather demanding task to accurately
control the stirring.

\begin{figure}
\begin{center}
\includegraphics[width=8cm]{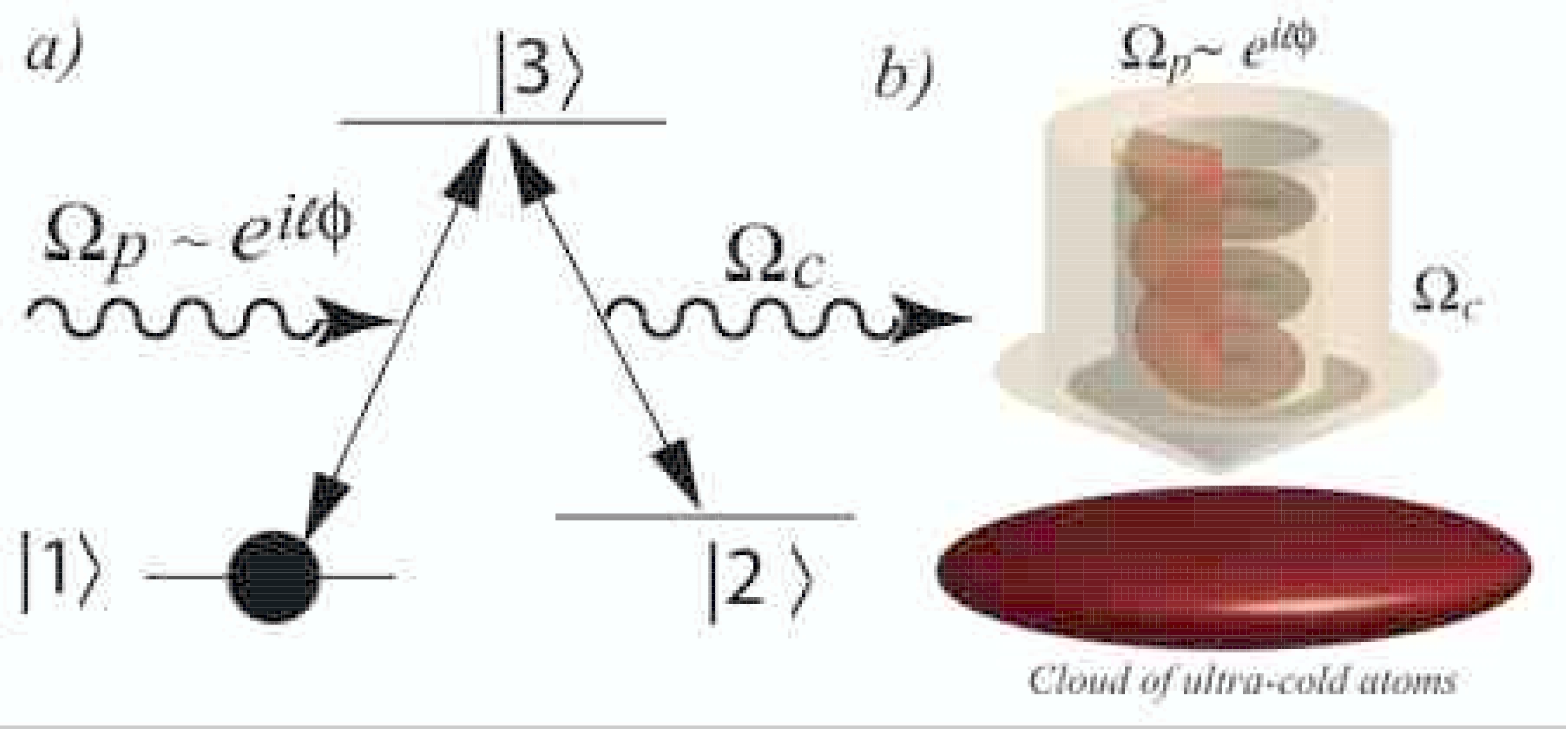}
\end{center}
\caption{a) The level scheme for the Electromagnetically Induced Transparency
with the probe beam and control beams characterized by the Rabi frequencies $\Omega_p$ and $\Omega_c$. b) The experimental
setup with the two copropagating light beams and the cloud of cold atoms.}
\label{exp}
\end{figure}

In recent papers \cite{juzeliunas04,longpaper05} we have shown how to create an effective magnetic field without stirring, using two light beams (to be referred to as
the control and probe beams) where at least one of them carries an orbital
angular momentum (OAM). The effective magnetic field stems from the interaction
of the laser beams with a medium of three level atoms in the
electromagnetically induced transparency (EIT) configuration, see Fig. \ref{exp} .
There is a significant advantage in creating the effective magnetic field using
light, since the key to the form of the magnetic field lies in the phase and
intensity of the light, concepts which with recent holographic
techniques can be taylored to a remarkable degree nowadays. We are therefore
now in a situation where we can choose different types of vector potentials and
study their influences on quantum gases both for fermions and bosons.

In the present paper we investigate the physical properties of a degenerate
two-dimensional Fermi gas of atoms in the presence of an effective
magnetic field. We start by introducing the concept of a vector potential and
the light-matter coupling. Subsequently we consider a trapped degenerate Fermi
gas and calculate the effects of a strong magnetic field. Finally we conclude by discussing the experimental implications and some future prospects.

\section{The model}

Let us consider a neutral cloud of three level atoms interacting with two
incident beams of light: a probe beam containing an orbital angular momentum
and a uniform control beam. The atoms are characterised by two hyperfine
ground states $1$ and $2$ and an excited electronic state $3$, see Fig.~\ref{exp}. The
two laser beams drive the atoms to the dark state,
$|D\rangle\sim\Omega_c |1\rangle -\Omega_p |2\rangle$, representing a coherent superposition
of the two hyperfine ground states $1$ and $2$. The corresponding equation for the wave function $\Phi_D$ representing the translational motion of the dark-state atoms  is derived in Ref. \cite{longpaper05},
\begin{equation}
i\hbar\partial_t\Phi_D=\frac{1}{2m}\left[i\hbar\mathbf{\nabla}+\mathbf{A}_{
\mathrm{eff}}\right]^2\Phi_D+V_{\mathrm{eff}}(\mathbf{r})\Phi_D,
\label{Eq-psi-1}
\end{equation}
where 
\begin{equation}
\mathbf{A}_{\mathrm{eff}}=\frac{i\hbar}{2}\frac{\zeta^*\mathbf{\nabla}\zeta
-\zeta\mathbf{\nabla}\zeta^*}{1+|\zeta^2|}\equiv-\hbar\frac{|\zeta|^2}{1
+|\zeta^2|}\mathbf{\nabla}S
\label{A-eff-rez}
\end{equation}
and 
\begin{equation}
V_{\mathrm{eff}}(\mathbf{r})=V_{\mathrm{ext}}(\mathbf{r})+\frac{\hbar^2}{
2m}\frac{|\zeta|^2(\nabla
S)^2+(\nabla|\zeta|)^2}{\left(1+|\zeta|^2\right)^2},
\label{V-eff-rez}
\end{equation}
are the \textit{effective vector potential} and \textit{effective trapping
potential}.   The external trapping potential for the dark-state atoms is given by 
\begin{equation}
V_{\mathrm{ext}}(\mathbf{r})=\frac{V_1(\mathbf{r})+|\zeta|^2(V_2(\mathbf{r})
+\epsilon_{21})}{1+|\zeta|^2}
\label{vext}
\end{equation}
with $V_j$ being the
trapping potential for the atoms in the hyperfine state $j$ ($j=1,2$), and
\begin{equation}
\epsilon_{21}=\hbar \left( \omega_{2}-\omega _{1}+\omega _{c}-\omega _{p}\right)
\end{equation}
is the energy of the two-photon detuning with $\hbar\omega_i$ the energies of the hyperfine states. The dimensionless
function $\zeta=\Omega_p/\Omega_c\equiv e^{iS}|\zeta|$ denotes the ratio
between the Rabi frequencies for the probe and control beams, where
$S=(\mathbf{k}_p-\mathbf{k}_c)\cdot\mathbf{r}+\ell\phi$ is the relative phase
of the two beams, $\mathbf{k}_p$ and $\mathbf{k}_c$ are the
wave-vectors, $\ell$ the winding number of the probe beam, and $\phi$ the
azimuthal angle.

In this way, the incident light field will act as a vector
potential as in Eq.~(\ref{Eq-psi-1}). The appearance of $\mathbf{A}_{\mathrm{eff}}$ is a manifestation 
of the Berry connection which is encountered in
many different areas of physics \cite{jack03,Sun90,Dum96}.
If we choose the control and probe beams co-propagating, with the probe beam
having an orbital angular momentum $\hbar\ell$ per photon and the
intensity of the form
\begin{equation}
|\zeta|^2=\frac{\alpha_0(r/R)^2}{1-\alpha_0(r/R)^2}\,,
\label{zeta-exploding}
\end{equation}
we obtain a uniform magnetic field in the $z$-direction 
$\mathbf{B}=-2\hbar\alpha_0\ell R^{-2}\hat{\mathbf{e}}_z$, with  
$\mathbf{A}_{\mathrm{eff}}=-\hbar\alpha_0\ell
rR^{-2}\hat{\mathbf{e}}_{\phi}$ and $\alpha_0<1$ a dimensionless parameter.
In the following we will choose the harmonic trapping potentials $V_1({\bf r})$ and 
$V_2({\bf r})$ such that $V_{eff}({\bf r})=0$ for $r<R$. In Ref. \cite{longpaper05} it is illustrated how this can be achieved using external potentials which are approximately harmonic, resulting in $V_{eff}$ being close to zero over a large region. In addition we assume here a steep barrier at $r=R$. Such barriers have recently been experimentally demonstrated using optical potentials \cite{box_bec}. 

The atoms can be safely considered noninteracting, since for spin polarised fermions
only weak p-wave scattering is present 
\cite{butts97,demarco99,mewes00,juzeliunas01}. The corresponding single particle states 
describing the trapped fermions are governed by the equation
\begin{eqnarray}
&&\bigg\{\frac{\hbar^2}{2m}\left[-\nabla^2+\left(\frac{\ell\alpha_0}{R^2}\right)^2r^2
+2i\left(\frac{\ell\alpha_0}{R^2}\right)\partial_{\phi}\right]\nonumber \\ &&+V_{eff}(r)\bigg\}    \Phi_D=E\Phi_D.
\label{sch1}
\end{eqnarray}

\section{Landau levels}

Let us assume we have a two-dimensional Fermi gas with the atomic motion confined to the $xy$ plane. 
After rescaling the radial coordinate, $r=xR$, and using the ansatz
$\Phi_D=\xi(x)e^{iq\phi}$ we obtain the solution in the form of a confluent
hypergeometric function 
\begin{align}
\xi(x) & =x^{|q|}e^{-\frac{|\ell|\alpha_{0}}{2}x^{2}}\nonumber \\ &
{}\times{}_{1}F_{1}\left[\frac{1+|q|}{2}-\left(\frac{\epsilon}{4|\ell|\alpha_{0}}+\frac{q}{2}\right),|q|+1;|\ell|\alpha_{0}x^{2}\right]\label{sol}
\end{align}
where $\epsilon=(E-E_z)\frac{2mR^2}{\hbar^2}$ and $E_z$ is the transversal
groundstate energy.  As such, Eq. (\ref{sol}) is
rather intractable. We can, however, obtain analytical expressions for the
eigenvalues in the limit $|\ell|\alpha_0\gg1$, where the energies are of the
form 
\begin{equation}
\epsilon_{n,q}=2|\ell|\alpha_0(2n+|q|-q+1)
\label{strong}
\end{equation}
with $n=0,1,2,...$ and $q=...-2,-1,0,1,2...$. This is indeed the Landau result. 
The Landau system is strictly defined for an untrapped gas, but for 
$|\ell|\alpha_0\gg1$, the boundary at $r=R$ has little effect on the energies \cite{lanlifqm}. 
Note that the energy levels in Eq. (\ref{strong}) are highly degenerate and are spaced by 
$4\ell\alpha_0$. These levels are equivalent to the Landau levels of the charged system.
The eigenstates for the Landau states are of the form 
\begin{equation}
\xi(x)=e^{iq\phi}x^{|q|}e^{-|\ell|\alpha_0x^2/2}L_n^{|q|}(|\ell|\alpha_0x^2)
\label{states}
\end{equation}
where $L_n^{|q|}(|\ell|\alpha_0x^2)$ is the Laguerre polynomials. 

Using the corresponding magnetic length $\ell_c=\frac{R}{\sqrt{2\alpha_0\ell}}$
the magnetic flux becomes $N_{\phi}=R^2/\ell_c^2=2\alpha_0\ell$.
The Fermi gas is therefore described by the completely filled lowest Landau
level if the criterion $N=N_{\phi}$ is fulfilled where $N$ is the number of
atoms. On the other hand, it should be noted, that the value of $N_{\phi}$ and the degeneracy is also limited by the fact that we have a finite trap.

It is at this point important to realise that the winding number $l$ of the light beams can with present technologies be of the order of a few
hundred \cite{high_ell_johannes,high_ell_curtis}, whereas the parameter $\alpha_0$ is smaller than $1$. A high optical orbital angular momentum is achieved by creating a highly charged optical vortex. There are many different techniques to create optical vortices. Highly charged optical vortices are typically made using spatial light modulators which act as a phase hologram, where the grating can be programmed to achieve the required phase of the light beam. This method is mainly limited by the pixel density in the hologram.  

\section{Characteristics of the filled Landau level gas}

We will illustrate the completely filled Landau level gas by looking at the static and dynamical 
phenomena arising due to the effective magnetic field. One important question in this respect is, how to distuingish between a gas described by the completely filled lowest Landau level and a gas which occupies more than one Landau levels? In the atomic case the situation is slightly more subtle compered to the normal quantum Hall situation with electrons since with non-interacting atoms the concept of resistivity is not necessarily a useful and well defined one.

We start by calculating the single
particle mass current. In order to do this we have to use the correct form of
the current operator \cite{me02_1} which now takes the form
\begin{equation}
J_k(x)=-\frac{i}{2}[\Psi^{\dag}(D_k\Psi)-(D_k\Psi)^{\dag}\Psi]
\label{current}
\end{equation}
where $D_k=\partial_k-iA_{eff}^k/\hbar$. Using the eigenstates in
Eq.~(\ref{states}) and Eq.~(\ref{current}) we obtain the current density 
\begin{equation}
\mathbf{J}^q(x)=\xi_q^*\mathbf{J}\xi_q=C_q^2 x^{2q-1}e^{-\alpha_0\ell
x^2}(\frac{q}{\alpha_0\ell}-x^2)\hat{\mathbf{e}}_{\phi}.
\end{equation}
where $C_q$ is a normalisation constant. The current is clearly zero at the
central distance $r_{\ell}=\sqrt{|q|/\alpha_0\ell}$ and flows in opposite directions on either side of $r_{\ell}$. The total current is consequently going to be zero. 

\begin{figure}
\begin{center}
\includegraphics[width=8.5cm]{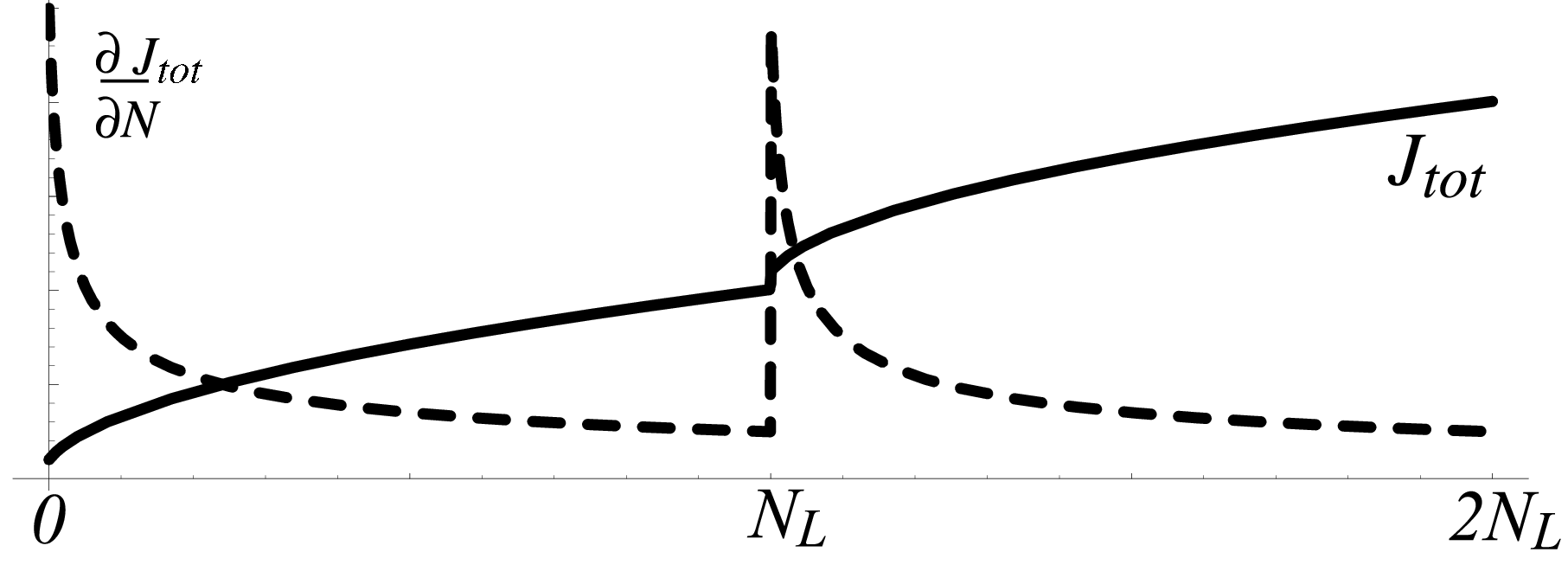}
\end{center}
\caption{The current as a function of particle number $N$ and its derivative
clearly show the transition between the Landau levels at $N=N_L$.}
\label{fig:current}
\end{figure}

In the spirit of the integer quantum Hall effect in a Corbino geometry we may ask ourselves, what happens if we add a potential linear in $r$ of the form
$V(r)=\beta r$? If $mR^2 \beta\ell_c/\hbar^2(\ell\alpha_0)^2 \ll 1$
the solutions of Eq. (\ref{sch1}) can be approximated by shifting the
solutions in Eq.(\ref{states}) by the factor $\frac{\beta}{2\alpha_0\ell}$. The
resulting single particle current density then takes the form 
\begin{equation}
\mathbf{J}^q(x) =C_q^2(x-\frac{1}{2}\frac{
\beta}{\alpha_0\ell})^{2q}e^{-\alpha_0\ell(x-\frac{1}{2}\frac{\beta}{
\alpha_0\ell})^2}\frac{1}{x}(\frac{q}{\alpha_0\ell}-x^2)\hat{\mathbf{e}}_{\phi}
\end{equation}
where the current is no longer zero as in the previous case. 
The total current per particle in $\phi$-direction and in the lowest Landau level, becomes $\mathbf{J}_{tot}^q=\int dx\mathbf{J}^q(x)=\beta\sqrt{\alpha_0\ell} \Gamma[q+1/2]/(4q!)\approx \beta\sqrt{\alpha_0\ell}/(4\sqrt{q})$. This also shows
that the velocity goes like $1/r$ since the single particle state is centred at
$r=r_{\ell}=\sqrt{2q}R/\sqrt{\alpha_0\ell}$, hence the flow is irrotational.
On the level of our approximations the current does not depend on which Landau level is occupied. Therefore the current depends nontrivially on the particle
number which manifests itself as a jump in the derivative 
with respect to particle number or magnetic field. In Fig.~\ref{fig:current} we show the total
current, $\mathbf{J}_{\mathrm{tot}}=\sum_{q=0}^{N-1} \mathbf{J}_{tot}^q =\sqrt{\alpha_0\ell}\beta \Gamma[N+1/2]/(2(N-1)!)\approx \beta\sqrt{\alpha_0\ell N}/2$, as a function of particle number $N$. 

\begin{figure}[h]
\begin{center}
\includegraphics[width=8.5cm]{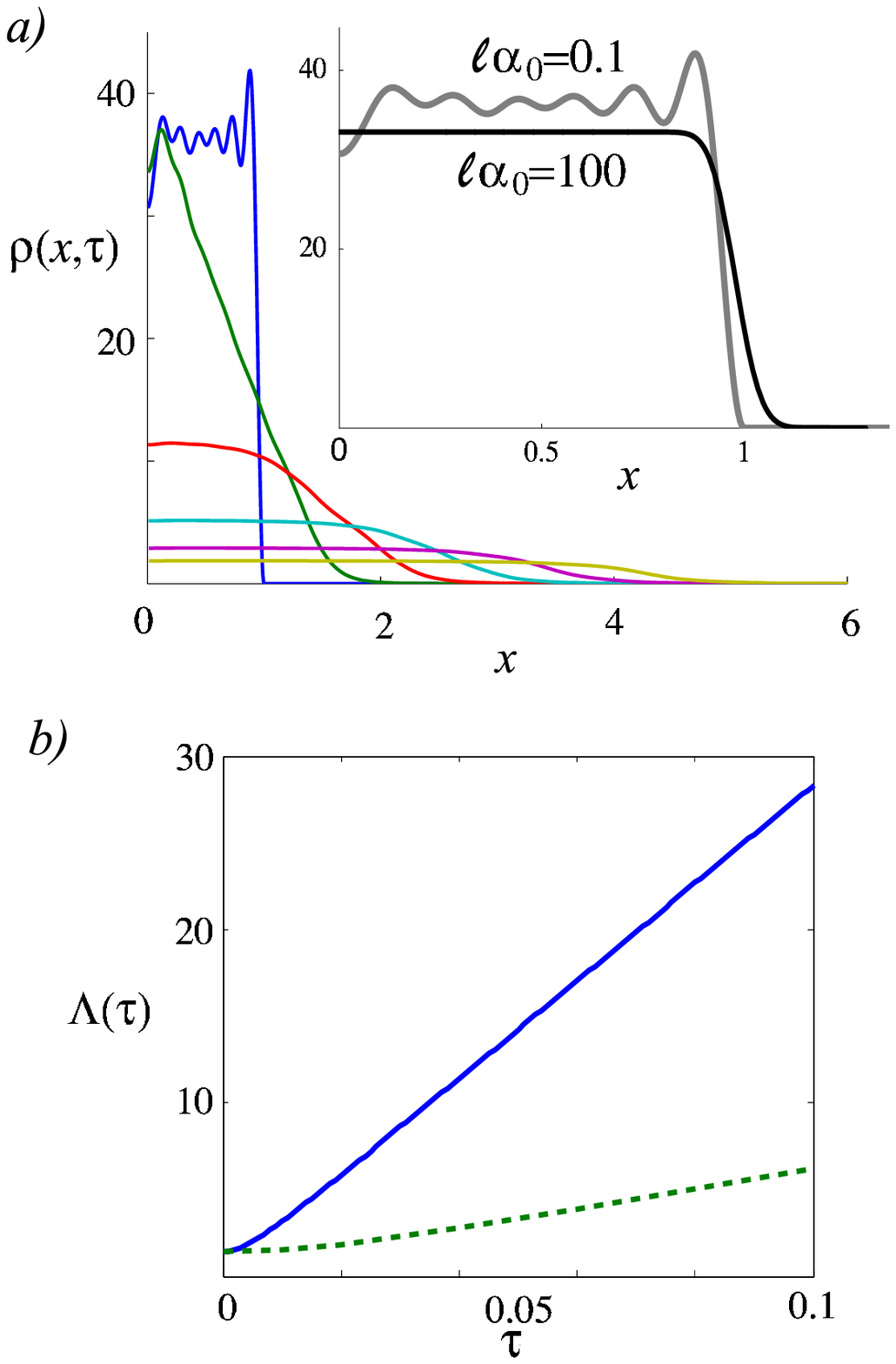}
\end{center}
\caption{(Color online) a) The non-self similar expansion of the cloud of atoms
corresponding to a weak magnetic field where many Landau levels are occupied
($\alpha_0\ell=0.1$) and $V_{eff}=0$. The different curves correspond to the times $\tau=0, 0.02, 0.04, 0.06, 0.08, 0.10$ where $\tau$ is in units of $2mR^2/\hbar$. The inset shows a comparison between the densities at $\tau=0$ for $\ell\alpha_0=0.1$ which occupies many Landau levels and the
cloud of atoms described by the single completely filled lowest Landau state, $N=\ell\alpha_0=100$ and $V_{eff}(x)=0$ for $x<1$. In both figures the total number of
particles were $N=100$. b) The diameter, $\Lambda(\tau)=2\sqrt{\int
d\mathbf{r}r^2\rho(r)}/\sqrt{N}$, for the completely filled Landau state
is seen to expand much more rapidly than the state corresponding to a weak
magnetic field ($\alpha_0\ell=0.1$).} \label{expansion}
\end{figure}

The physics of the lowest completely filled Landau level is in itself an
interesting concept and shows some rather intriguing scenarios both from a
fundamental and experimental point of view \cite{hall1}. 
One of the most important questions concerning the atomic quantum hall state is
what to measure? For a gas described by the completely filled Landau level, the density is
going to be homogeneous, since the density is effectively built up by shifted
Gaussians corresponding to the different angular momenta as can be seen in
Eq.~(\ref{sol}). This also means we have added an external
potential in order to have $V_{eff}=0$. This is not necessary but makes the
situation simpler and more intuitive. Consequently, if there is no probe beam
carrying orbital angular momentum propagating through the gas, the atoms will
be subject to an external harmonic trap and hence will show a density profile
quadratic in $r$, clearly different than the filled Landau state, see inset in Fig.~(\ref{expansion}).
From an experimentalist's point of view a direct observation of the density of
the cloud in the trap is not necessarily the most convenient way of observing
the atoms. Another possibility is to consider the free expansion of the cloud.

Since the trapped fermions can be considered noninteracting we
can calculate the dynamics of the freely expanding cloud using the single particle propagator, 
\begin{equation}
K(r,r',\phi,\phi';t)=\frac{1}{i4\pi\tau}e^{\frac{i}{4\tau}(r^2+r'^2-2rr'\cos(\phi-\phi'))}.
\label{prop}
\end{equation}
where $\tau$ is the rescaled time $t=\frac{2mR^2}{\hbar}\tau$ and $r$ is in
units of $R$. After a straightforward integration we obtain the dynamics of the
freely expanding cloud using the states in Eq.~(\ref{sol}), see Fig.~(\ref{expansion}). The
dynamics of the cloud described by the single completely filled Landau level is self similar and is  captured by the scaling
parameter $\sigma_L=\sqrt{1+4\tau^2(\alpha_0\ell)^2}$.  In Fig.~(\ref{expansion}) we show the
mean width defined as $\Lambda(\tau)=2\sqrt{\int
d\mathbf{r}r^2\rho(r)}/\sqrt{N}$ which can readily be calculated using the
states in Eq.~(\ref{sol}). The lowest
completely filled Landau level is found to expand as 
$\Lambda(\tau)=\sigma_L(\tau)\sqrt{\frac{2(N+1)}{\ell\alpha_0}}$ \cite{comment}.

Clearly the expansion of a single completely filled lowest Landau level
compared to the situation with also the second Landau level filled, is not
going to be much different, since the expansion is still self similar, apart 
from contributions from the edge states. The
density of the trapped cloud is going to be homogeneous and will remain so when
expanding. This is seen from the energy spectrum in Eq.~(\ref{strong}). For an
energy corresponding to the second Landau level we can have $n=1$ and $q\geq0$
but also $n=0$ and $q=-1$. But this correction is only of the order of a single
particle and will not be measurable, hence the expansion dynamics will be the
same as the lowest Landau level. There will, however, be a significant
difference if we consider the two extreme situations with a Fermi gas described
as a completely filled lowest Landau level and the situation when the effective
magnetic field is weak. If the magnetic field is weak the trapped Fermi gas is
well described by spherical Bessel functions corresponding to the
eigenfunctions of a cylindrically trapped Fermi gas. The expansion dynamics can
still be calculated with the free particle propagator in Eq.~(\ref{prop}). The
important difference is evident in the expansion which is no longer going to be
self similar. Figure (\ref{expansion}) shows the mean width of the cloud as function of time
compared to the completely filled lowest Landau level. The broken
self-similarity is most clearly seen in a series of snap-shots of the density compared to
the density of the completely filled Landau level. The lowest
Landau level states expand much faster than states belonging to higher Landau
levels corresponding to the weak magnetic field case. This is easily understood since the lowest Landau 
level states are pure angular momenta states with no radial excitation, whereas the gas with a weak magnetic field contains 
states with radial excitations but lower angular momenta, hence lower kinetic energy.

\section{Conclusions}

Throughout this paper we have considered a two-dimensional trapped Fermi gas.
It is important to remember that a two-dimensional Fermi gas poses some rather
strict conditions on the external trap configuration. The two-dimensionality is
preserved if the Fermi energy is lower than the relevant transversal
groundstate energy. The relevant energy scale is here the effective cyclotron frequency
which for typical radii of the cloud (a few tens of microns) can be of the order of 
a hundred Hz, hence the transversal trap frequency needs to be significantly stronger
than this. Such traps are indeed readily available \cite{ketterle_lowd}. The effective magnetic field
relies on the stability of the dark state. As discussed in Ref. \cite{longpaper05}, with typical experimental parameters, the dark state will be stable for times significantly longer than the normal life 
time for a trapped cloud, making the effective magnetic field created by optical orbital angular momentum a feasible technique for achieving strong magnetic fields. 
Clearly from an experimental point of view, the biggest
challenge is found in the detection of the mass current in the cloud. There are however powerful 
techniques based on slow light propagation and the dragging of the light \cite{me02_1,juz+mas+fle03} which will identify a mass current in the cloud {\it in situ}. Another possibility is to measure the shape
oscillations which should be affected by the current.

In this paper we have investigated the concept of completely filled Landau
levels in trapped Fermi gases. The effective magnetic field was created using
light with orbital angular momentum. The recent advances in creating exotic
light beams where both phase and intensity can be manipulated, allows us to
consider many different forms of the effective magnetic field. We have
restricted ourselves to the ``textbook'' scenario with a
homogeneous effective magnetic field.  It is important to note
here that this scenario is different from other techniques where
filled Landau levels are considered. In the case of bosons in a harmonic trap, the
gas is stirred to create many vortices resulting in an angular momentum which
would correspond to the trap groundstate energy. In our case the situation is
much simpler. Only a static external trap needs to be added which matches the
corresponding cyclotron frequency. If the frequency of the added harmonic trap
does not match the cyclotron frequency, the important degeneracy will be
lifted, but as long as
\begin{equation}
N=q_{max}<\frac{2}{1-1/\sqrt{1+(\Delta\omega/\omega_c)^2}}\approx4(\frac{
\omega_c}{\Delta\omega})^2
\end{equation}
where $\omega_c=\hbar\alpha_0\ell/(mR^2)$ and $\Delta\omega$ is the
preferably small deviation, the atoms which fill the lowest Landau level will
not mix with the higher levels.

\begin{acknowledgments}
This work was supported by the Royal Society of
Edinburgh, the Alexander von Humboldt foundation and the Marie-Curie
Trainings-site at the University of Kaiserslautern. 
Helpful discussions with J.~Courtial, S. Giovanazzi, M.~Lindberg and S.~Stenholm are also
greatly acknowledged. 
\end{acknowledgments}

\begin{thebibliography}{30}
\expandafter\ifx\csname natexlab\endcsname\relax\def\natexlab#1{#1}\fi
\expandafter\ifx\csname bibnamefont\endcsname\relax
  \def\bibnamefont#1{#1}\fi
\expandafter\ifx\csname bibfnamefont\endcsname\relax
  \def\bibfnamefont#1{#1}\fi
\expandafter\ifx\csname citenamefont\endcsname\relax
  \def\citenamefont#1{#1}\fi
\expandafter\ifx\csname url\endcsname\relax
  \def\url#1{\texttt{#1}}\fi
\expandafter\ifx\csname urlprefix\endcsname\relax\def\urlprefix{URL }\fi
\providecommand{\bibinfo}[2]{#2}
\providecommand{\eprint}[2][]{\url{#2}}

\bibitem[{\citenamefont{Butts and Rokhsar}(1997)}]{butts97}
\bibinfo{author}{\bibfnamefont{D.~A.} \bibnamefont{Butts}} \bibnamefont{and}
  \bibinfo{author}{\bibfnamefont{D.~S.} \bibnamefont{Rokhsar}},
  \bibinfo{journal}{Phys. Rev. A} \textbf{\bibinfo{volume}{55}},
  \bibinfo{pages}{4346} (\bibinfo{year}{1997}).

\bibitem[{\citenamefont{DeMarco and Jin}(1999)}]{demarco99}
\bibinfo{author}{\bibfnamefont{B.}~\bibnamefont{DeMarco}} \bibnamefont{and}
  \bibinfo{author}{\bibfnamefont{D.~S.} \bibnamefont{Jin}},
  \bibinfo{journal}{Science} \textbf{\bibinfo{volume}{285}},
  \bibinfo{pages}{1703} (\bibinfo{year}{1999}).

\bibitem[{\citenamefont{Mewes et~al.}(2000)\citenamefont{Mewes, Ferrari,
  Schreck, Sinatra, and Salomon}}]{mewes00}
\bibinfo{author}{\bibfnamefont{M.-O.} \bibnamefont{Mewes}},
  \bibinfo{author}{\bibfnamefont{G.}~\bibnamefont{Ferrari}},
  \bibinfo{author}{\bibfnamefont{F.}~\bibnamefont{Schreck}},
  \bibinfo{author}{\bibfnamefont{A.}~\bibnamefont{Sinatra}}, \bibnamefont{and}
  \bibinfo{author}{\bibfnamefont{C.}~\bibnamefont{Salomon}},
  \bibinfo{journal}{Phys. Rev. A} \textbf{\bibinfo{volume}{61}},
  \bibinfo{pages}{011403(R)} (\bibinfo{year}{2000}).

\bibitem[{\citenamefont{Pitaevskii and Stringari}(2003)}]{bec_stri}
\bibinfo{author}{\bibfnamefont{L.}~\bibnamefont{Pitaevskii}} \bibnamefont{and}
  \bibinfo{author}{\bibfnamefont{S.}~\bibnamefont{Stringari}},
  \emph{\bibinfo{title}{Bose-Einstein Condensation}}
  (\bibinfo{publisher}{Clarendon Press, Oxford}, \bibinfo{year}{2003}).

\bibitem[{\citenamefont{Jaksch and Zoller}(2003)}]{jaksch03}
\bibinfo{author}{\bibfnamefont{D.}~\bibnamefont{Jaksch}} \bibnamefont{and}
  \bibinfo{author}{\bibfnamefont{P.}~\bibnamefont{Zoller}},
  \bibinfo{journal}{New J. Phys.} \textbf{\bibinfo{volume}{5}},
  \bibinfo{pages}{56} (\bibinfo{year}{2003}).

\bibitem[{\citenamefont{Sorensen et~al.}()\citenamefont{Sorensen, Demler, and
  Lukin}}]{anders04}
\bibinfo{author}{\bibfnamefont{A.}~\bibnamefont{Sorensen}},
  \bibinfo{author}{\bibfnamefont{E.}~\bibnamefont{Demler}}, \bibnamefont{and}
  \bibinfo{author}{\bibfnamefont{M.}~\bibnamefont{Lukin}},
  \eprint{cond-mat/0405079}.

\bibitem[{\citenamefont{Mueller}(2004)}]{mueller04}
\bibinfo{author}{\bibfnamefont{E.~J.} \bibnamefont{Mueller}},
  \bibinfo{journal}{Phys. Rev. A} \textbf{\bibinfo{volume}{70}},
  \bibinfo{pages}{041603(R)} (\bibinfo{year}{2004}).

\bibitem[{\citenamefont{Osterloh et~al.}(2005)\citenamefont{Osterloh, Baig,
  Santos, Zoller, and Lewenstein}}]{levenstein-non-abelian-preprint05}
\bibinfo{author}{\bibfnamefont{K.}~\bibnamefont{Osterloh}},
  \bibinfo{author}{\bibfnamefont{M.}~\bibnamefont{Baig}},
  \bibinfo{author}{\bibfnamefont{L.}~\bibnamefont{Santos}},
  \bibinfo{author}{\bibfnamefont{P.}~\bibnamefont{Zoller}}, \bibnamefont{and}
  \bibinfo{author}{\bibfnamefont{M.}~\bibnamefont{Lewenstein}},
  \bibinfo{journal}{Phys. Rev. Lett} \textbf{\bibinfo{volume}{95}},
  \bibinfo{pages}{010403} (\bibinfo{year}{2005}).

\bibitem[{\citenamefont{Bretin et~al.}(2004)\citenamefont{Bretin, Stock,
  Seurin, and Dalibard}}]{bretin04}
\bibinfo{author}{\bibfnamefont{V.}~\bibnamefont{Bretin}},
  \bibinfo{author}{\bibfnamefont{S.}~\bibnamefont{Stock}},
  \bibinfo{author}{\bibfnamefont{Y.}~\bibnamefont{Seurin}}, \bibnamefont{and}
  \bibinfo{author}{\bibfnamefont{J.}~\bibnamefont{Dalibard}},
  \bibinfo{journal}{Phys. Rev. Lett} \textbf{\bibinfo{volume}{92}},
  \bibinfo{pages}{050403} (\bibinfo{year}{2004}).

\bibitem[{\citenamefont{Schweikhard et~al.}(2004)\citenamefont{Schweikhard,
  Coddington, Engels, Mogendorff, and Cornell}}]{schweikhard04}
\bibinfo{author}{\bibfnamefont{V.}~\bibnamefont{Schweikhard}},
  \bibinfo{author}{\bibfnamefont{I.}~\bibnamefont{Coddington}},
  \bibinfo{author}{\bibfnamefont{P.}~\bibnamefont{Engels}},
  \bibinfo{author}{\bibfnamefont{V.~P.} \bibnamefont{Mogendorff}},
  \bibnamefont{and} \bibinfo{author}{\bibfnamefont{E.~A.}
  \bibnamefont{Cornell}}, \bibinfo{journal}{Phys. Rev. Lett}
  \textbf{\bibinfo{volume}{92}}, \bibinfo{pages}{040404}
  (\bibinfo{year}{2004}).

\bibitem[{\citenamefont{Baym and Pethick}(2004)}]{Baym03}
\bibinfo{author}{\bibfnamefont{G.}~\bibnamefont{Baym}} \bibnamefont{and}
  \bibinfo{author}{\bibfnamefont{C.~J.} \bibnamefont{Pethick}},
  \bibinfo{journal}{Phys. Rev. A} \textbf{\bibinfo{volume}{69}},
  \bibinfo{pages}{043619} (\bibinfo{year}{2004}).

\bibitem[{\citenamefont{Fischer and Baym}(2003)}]{fischer03}
\bibinfo{author}{\bibfnamefont{U.~R.} \bibnamefont{Fischer}} \bibnamefont{and}
  \bibinfo{author}{\bibfnamefont{G.}~\bibnamefont{Baym}},
  \bibinfo{journal}{Phys. Rev. Lett} \textbf{\bibinfo{volume}{90}},
  \bibinfo{pages}{140402} (\bibinfo{year}{2003}).

\bibitem[{\citenamefont{Cooper et~al.}(2001)\citenamefont{Cooper, Wilkin, and
  Gunn}}]{cooper01}
\bibinfo{author}{\bibfnamefont{N.~R.} \bibnamefont{Cooper}},
  \bibinfo{author}{\bibfnamefont{N.~K.} \bibnamefont{Wilkin}},
  \bibnamefont{and} \bibinfo{author}{\bibfnamefont{J.~M.~F.}
  \bibnamefont{Gunn}}, \bibinfo{journal}{Phys. Rev. Lett}
  \textbf{\bibinfo{volume}{87}}, \bibinfo{pages}{120405}
  (\bibinfo{year}{2001}).

\bibitem[{\citenamefont{Ho}(2001)}]{ho01}
\bibinfo{author}{\bibfnamefont{T.-L.} \bibnamefont{Ho}},
  \bibinfo{journal}{Phys. Rev. Lett} \textbf{\bibinfo{volume}{87}},
  \bibinfo{pages}{060403} (\bibinfo{year}{2001}).

\bibitem[{\citenamefont{Baranov et~al.}()\citenamefont{Baranov, Osterloh, and
  Lewenstein}}]{baranov04}
\bibinfo{author}{\bibfnamefont{M.~A.} \bibnamefont{Baranov}},
  \bibinfo{author}{\bibfnamefont{K.}~\bibnamefont{Osterloh}}, \bibnamefont{and}
  \bibinfo{author}{\bibfnamefont{M.}~\bibnamefont{Lewenstein}},
  \eprint{cond-mat/0404329}.

\bibitem[{\citenamefont{Juzeli{\=u}nas and {\"O}hberg}(2004)}]{juzeliunas04}
\bibinfo{author}{\bibfnamefont{G.}~\bibnamefont{Juzeli{\=u}nas}}
  \bibnamefont{and}
  \bibinfo{author}{\bibfnamefont{P.}~\bibnamefont{{\"O}hberg}},
  \bibinfo{journal}{Phys. Rev. Lett} \textbf{\bibinfo{volume}{93}},
  \bibinfo{pages}{033602} (\bibinfo{year}{2004}).

\bibitem[{\citenamefont{Juzeli{\=u}nas
  et~al.}(2005)\citenamefont{Juzeli{\=u}nas, {\"O}hberg, Ruseckas, and
  Klein}}]{longpaper05}
\bibinfo{author}{\bibfnamefont{G.}~\bibnamefont{Juzeli{\=u}nas}},
  \bibinfo{author}{\bibfnamefont{P.}~\bibnamefont{{\"O}hberg}},
  \bibinfo{author}{\bibfnamefont{J.}~\bibnamefont{Ruseckas}}, \bibnamefont{and}
  \bibinfo{author}{\bibfnamefont{A.}~\bibnamefont{Klein}},
  \bibinfo{journal}{Phys. Rev. A} \textbf{\bibinfo{volume}{71}},
  \bibinfo{pages}{053614} (\bibinfo{year}{2005}).

\bibitem[{\citenamefont{Jackiw}(1988)}]{jack03}
\bibinfo{author}{\bibfnamefont{R.}~\bibnamefont{Jackiw}},
  \bibinfo{journal}{Comments At. Mol. Phys.} \textbf{\bibinfo{volume}{21}},
  \bibinfo{pages}{71} (\bibinfo{year}{1988}).

\bibitem[{\citenamefont{Sun and Ge}(1990)}]{Sun90}
\bibinfo{author}{\bibfnamefont{C.-P.} \bibnamefont{Sun}} \bibnamefont{and}
  \bibinfo{author}{\bibfnamefont{M.-L.} \bibnamefont{Ge}},
  \bibinfo{journal}{Phys. Rev. D} \textbf{\bibinfo{volume}{41}},
  \bibinfo{pages}{1349} (\bibinfo{year}{1990}).

\bibitem[{\citenamefont{Dum and Olshanii}(1996)}]{Dum96}
\bibinfo{author}{\bibfnamefont{R.}~\bibnamefont{Dum}} \bibnamefont{and}
  \bibinfo{author}{\bibfnamefont{M.}~\bibnamefont{Olshanii}},
  \bibinfo{journal}{Phys. Rev. Lett.} \textbf{\bibinfo{volume}{76}},
  \bibinfo{pages}{1788} (\bibinfo{year}{1996}).

\bibitem[{\citenamefont{Meyrath et~al.}()\citenamefont{Meyrath, Schreck,
  Hanssen, Chuu, and Raizen}}]{box_bec}
\bibinfo{author}{\bibfnamefont{T.}~\bibnamefont{Meyrath}},
  \bibinfo{author}{\bibfnamefont{F.}~\bibnamefont{Schreck}},
  \bibinfo{author}{\bibfnamefont{J.}~\bibnamefont{Hanssen}},
  \bibinfo{author}{\bibfnamefont{C.-S.} \bibnamefont{Chuu}}, \bibnamefont{and}
  \bibinfo{author}{\bibfnamefont{M.}~\bibnamefont{Raizen}},
  \eprint{cond-mat/0503590}.

\bibitem[{\citenamefont{Juzeli{\=u}nas and Ma{\v s}alas}(2001)}]{juzeliunas01}
\bibinfo{author}{\bibfnamefont{G.}~\bibnamefont{Juzeli{\=u}nas}}
  \bibnamefont{and} \bibinfo{author}{\bibfnamefont{M.}~\bibnamefont{Ma{\v
  s}alas}}, \bibinfo{journal}{Phys. Rev. A} \textbf{\bibinfo{volume}{63}},
  \bibinfo{pages}{061602(R)} (\bibinfo{year}{2001}).

\bibitem[{\citenamefont{Landau and Lifshitz}(2002)}]{lanlifqm}
\bibinfo{author}{\bibfnamefont{L.~D.} \bibnamefont{Landau}} \bibnamefont{and}
  \bibinfo{author}{\bibfnamefont{E.~M.} \bibnamefont{Lifshitz}},
  \emph{\bibinfo{title}{Quantum Mechanics}}
  (\bibinfo{publisher}{Butterworth-Heinemann, Oxford}, \bibinfo{year}{2002}).

\bibitem[{\citenamefont{Courtial et~al.}(1997)\citenamefont{Courtial, Dholakia,
  Allen, and Padgett}}]{high_ell_johannes}
\bibinfo{author}{\bibfnamefont{J.}~\bibnamefont{Courtial}},
  \bibinfo{author}{\bibfnamefont{K.}~\bibnamefont{Dholakia}},
  \bibinfo{author}{\bibfnamefont{L.}~\bibnamefont{Allen}}, \bibnamefont{and}
  \bibinfo{author}{\bibfnamefont{M.}~\bibnamefont{Padgett}},
  \bibinfo{journal}{Opt. Commun.} \textbf{\bibinfo{volume}{144}},
  \bibinfo{pages}{210} (\bibinfo{year}{1997}).

\bibitem[{\citenamefont{Curtis et~al.}(2002)\citenamefont{Curtis, Koss, and
  Grier}}]{high_ell_curtis}
\bibinfo{author}{\bibfnamefont{J.}~\bibnamefont{Curtis}},
  \bibinfo{author}{\bibfnamefont{B.}~\bibnamefont{Koss}}, \bibnamefont{and}
  \bibinfo{author}{\bibfnamefont{D.}~\bibnamefont{Grier}},
  \bibinfo{journal}{Optics Communications} \textbf{\bibinfo{volume}{207}},
  \bibinfo{pages}{169} (\bibinfo{year}{2002}).

\bibitem[{\citenamefont{{\"O}hberg}(2002)}]{me02_1}
\bibinfo{author}{\bibfnamefont{P.}~\bibnamefont{{\"O}hberg}},
  \bibinfo{journal}{Phys. Rev. A} \textbf{\bibinfo{volume}{66}},
  \bibinfo{pages}{021603(R)} (\bibinfo{year}{2002}).

\bibitem[{\citenamefont{Ezawa}(2000)}]{hall1}
\bibinfo{author}{\bibfnamefont{Z.~F.} \bibnamefont{Ezawa}},
  \emph{\bibinfo{title}{Quantum Hall Effects}} (\bibinfo{publisher}{World
  Scientific, Singapore}, \bibinfo{year}{2000}).

\bibitem[{com()}]{comment}Strictly speaking this is only true for an asymmetric dark state, $\alpha_0 \ll 1$. The weakly populated state $|2\rangle$ will expand faster than 
state $|1\rangle$ due to the acquired phase, causing a faint halo 
effect in the total density. This can be avoided if a STIRAP 
procedure is used before the expansion to transfer all 
population to state $|1\rangle$ by switching off the probe field.


\bibitem[{\citenamefont{G\"orlitz et~al.}(2001)\citenamefont{G\"orlitz, Vogels,
  Leanhardt, Raman, Gustavson, Abo-Shaeer, Chikkatur, Gupta, Inouye, Rosenband
  et~al.}}]{ketterle_lowd}
\bibinfo{author}{\bibfnamefont{A.}~\bibnamefont{G\"orlitz}},
  \bibinfo{author}{\bibfnamefont{J.~M.} \bibnamefont{Vogels}},
  \bibinfo{author}{\bibfnamefont{A.~E.} \bibnamefont{Leanhardt}},
  \bibinfo{author}{\bibfnamefont{C.}~\bibnamefont{Raman}},
  \bibinfo{author}{\bibfnamefont{T.~L.} \bibnamefont{Gustavson}},
  \bibinfo{author}{\bibfnamefont{J.~R.} \bibnamefont{Abo-Shaeer}},
  \bibinfo{author}{\bibfnamefont{A.~P.} \bibnamefont{Chikkatur}},
  \bibinfo{author}{\bibfnamefont{S.}~\bibnamefont{Gupta}},
  \bibinfo{author}{\bibfnamefont{S.}~\bibnamefont{Inouye}},
  \bibinfo{author}{\bibfnamefont{T.}~\bibnamefont{Rosenband}},
  \bibnamefont{et~al.}, \bibinfo{journal}{Phys. Rev. Lett.}
  \textbf{\bibinfo{volume}{87}}, \bibinfo{pages}{130402}
  (\bibinfo{year}{2001}).

\bibitem[{\citenamefont{Juzeli{\=u}nas
  et~al.}(2003)\citenamefont{Juzeli{\=u}nas, Ma{\v s}alas, and
  Fleischhauer}}]{juz+mas+fle03}
\bibinfo{author}{\bibfnamefont{G.}~\bibnamefont{Juzeli{\=u}nas}},
  \bibinfo{author}{\bibfnamefont{M.}~\bibnamefont{Ma{\v s}alas}},
  \bibnamefont{and}
  \bibinfo{author}{\bibfnamefont{M.}~\bibnamefont{Fleischhauer}},
  \bibinfo{journal}{Phys. Rev. A} \textbf{\bibinfo{volume}{67}},
  \bibinfo{pages}{023809} (\bibinfo{year}{2003}).

\end{thebibliography}

\end{document}